\documentstyle[pra,aps,epsfig,multicol]{revtex}

\begin{document}
\title{\Large\bf Structural, electronic and magnetic properties 
of heterofullerene ${\rm C}_{48}{\rm B}_{12}$}
\author{ Rui-Hua Xie$^{1}$,  Lasse Jensen$^{2}$, Garnett W. Bryant$^{1}$, 
Jijun Zhao$^{3}$, and Vedene H. Smith, Jr.$^{4}$} 
\address{$^{1}$National Institute of 
Standards and Technology, Gaithersburg, MD 20899-8423, USA\\ 
$^{2}$ Theoretical Chemistry, Materials Science Centre, 
Rijksuniversiteit Groningen, Nijenborgh 4,\\  9747 AG Groningen, The Netherlands\\ 
$^{3}$ Department of Physics and Astronomy, University of North Carolina, Chapel Hill, NC 27599, USA\\
$^{4}$ Department of Chemistry, Queen's University, Kingston,
ON K7L 3N6, Canada}
\date{\today}
\maketitle

\begin{abstract}
Bonding, electric (hyper)polarizability, vibrational and magnetic properties of 
heterofullerene ${\rm C}_{48}{\rm B}_{12}$ are studied by first-principles calculations.  
Infrared- and Raman-active vibrational frequencies of ${\rm C}_{48}{\rm B}_{12}$ 
are assigned. Eight $^{13}$C and 2 $^{11}$B nuclear magnetic resonance (NMR) 
spectral signals of ${\rm C}_{48}{\rm B}_{12}$  are characterized.  The average 
second-order hyperpolarizability of ${\rm C}_{48}{\rm B}_{12}$ is about 180\% larger 
than that of ${\rm C}_{60}$. Our results suggest that ${\rm C}_{48}{\rm B}_{12}$ is 
a  candidate for photonic and optical limiting applications because of the enhanced 
third-order optical nonlinearities. 
\end{abstract}

\begin{multicols}{2}

\begin{center}{\bf I. INTRODUCTION} \end{center}

In 1985, Kroto {\sl et al.}\cite{kroto85} proposed the existence of ${\rm C}_{60}$ 
clusters in their graphite laser vaporization experiment. This proposal was subsequently 
confirmed in 1990 when Kr\"{a}tschmer {\sl et al.}\cite{krat90} reported a method for 
the mass production of ${\rm C}_{60}$ in a carbon arc along with infra-red (IR) spectroscopic 
evidence for the C$_{60}$ carbon-structure. These pioneering works have  stimulated extensive 
research into fullerenes\cite{kroto93,book1},  a new form of pure carbon,  where an 
even number of three-coordinated ${\rm sp}^{2}$ carbon atoms arrange themselves into 12 
pentagonal faces and any number ($> 1$) of hexagonal faces. These carbon-cage molecules 
can crystallize into a variety of three-dimensional structures\cite{krat90} and  be 
doped in several different ways\cite{book1}:   endohedral doping, where the dopant is 
inside the fullerene cage; substitutional doping, where the dopant is on the fullerene 
cage; and exohedral doping, where the dopant is outside or 
between fullerene cages. It has been shown that doped fullerenes have remarkable 
structural, electronic, optical and magnetic properties \cite{book1,book4a,book4b}. 

In 1995, the heterofullerene ${\rm C}_{59}{\rm N}$ was formed efficiently in the gas phase 
during fast atom bombardment mass spectroscopy of a cluster-opened N-methoxyethoxy 
methyl ketolactam \cite{hummelen95}. The isolation and characterization of 
biazafullerenyl has opened a viable route for the preparation of  ${\rm C}_{59}{\rm N}$ 
and other heterofullerenes in solution, leading to a number of 
detailed theoretical and experimental studies  of 
${\rm C}_{59}{\rm N}$ and heterofullerenes\cite{book1,book4a,book4b}.   
In 1991, the Smalley group \cite{guo91} successfully synthesized boron-substituted fullerenes 
${\rm C}_{60-n}{\rm B}_{n}$ ($1\le n \le 6$). Very recently, Hultman {\sl et al.}
\cite{hultman01} have successfully synthesized aza-fullerenes 
${\rm C}_{60-n}{\rm N}_{n}$, formed by substituting carbon atoms in ${\rm C}_{60}$ with more than 
one nitrogen atom, and the existence of a stable ${\rm C}_{48}{\rm N}_{12}$ aza-fullerene 
\cite{hultman01,stafstrom,xie02a,mana02} was revealed.  Stimulated by the high stability of  
${\rm C}_{48}{\rm N}_{12}$, we have recently predicted that 
 ${\rm C}_{48}{\rm B}_{12}$\cite{xie02b} is also a stable heterofullerene and can 
be a promising component for molecular rectifiers, nanotube-based transistors and $p$-$n$ junctions. 

In this letter,  we  further study the bonding, Mulliken charges, 
electric (hyper)polarizability,  vibrational and magnetic properties of 
${\rm C}_{48}{\rm B}_{12}$. We  characterize $^{13}$C and $^{11}$B  NMR 
spectral lines of ${\rm C}_{48}{\rm B}_{12}$ and show how the  boron-substitutional 
doping modifies the infrared and Raman spectra of the pristine ${\rm C}_{60}$. 
We also find that ${\rm C}_{48}{\rm B}_{12}$ exhibits enhanced second-order hyperpolarizability (enhanced third-order 
optical nonlinearity) and can compete with ${\rm C}_{60}$ and aza-fullerene 
${\rm C}_{48}{\rm N}_{12}$ as a candidate for photonic and optical limiting applications 
(for examples, data processing, eye and sensor protection, all-optical switching, 
and optical limiting)\cite{book4b}. 

\begin{center}{\bf II. BONDING AND MULLIKEN CHARGE} \end{center}

The geometry of ${\rm C}_{48}{\rm B}_{12}$, shown in Fig.1, was fully optimized 
by using the Gaussian 98 program\cite{gaussian,nist}. We have used the B3LYP
\cite{becke93} hybrid density functional theory (DFT) method, which includes a mixture of Hartree-Fock (exact) exchange, 
Slater local exchange\cite{slater74}, Becke 88 non-local exchange\cite{becke88}, 
the VWN III local exchange-correlation functional \cite{vosko80} and the LYP 
correlation functional \cite{lyp88}, and a 6-31G(d) basis set. We consider of the form of 
C$_{48}$B$_{12}$ with such a dopant assignment: each pentagon has  one boron atom and two 
boron atoms preferentially sit in one hexagon, while other forms are possible. The 
symmetry of ${\rm C}_{48}{\rm B}_{12}$ is found to be a $C_{i}$ point group. 
The distances (or radii) ${\rm R}_{i}$ from the $i$th atom to the density 
center of the molecule are listed in Table I. We find  that  ${\rm C}_{48}{\rm B}_{12}$,  
similar to ${\rm C}_{48}{\rm N}_{12}$\cite{xie02a}, is an ellipsoid structure with 10 unique radii, 
while ${\rm C}_{60}$ has the same radius for each carbon atom (calculated 
$R_{i}$ = 0.35502 nm, in excellent agreement with experiment\cite{burgi92}). 

The calculated net Mulliken charges ${\rm Q}_{\rm i}$ of
carbon and boron atoms in ${\rm C}_{48}{\rm B}_{12}$
are  listed in Table I.  Like ${\rm C}_{48}{\rm N}_{12}$\cite{stafstrom,xie02a}, 
the heterofullerene ${\rm C}_{48}{\rm B}_{12}$ has two types of boron dopants 
 in the structure: one with net Mulliken charges 
${\rm Q}_{\rm i} = 0.1637\ e$
and the other with ${\rm Q}_{\rm i} = 0.1871\ e$. All carbon atoms in
${\rm C}_{48}{\rm B}_{12}$  have negative ${\rm Q}_{\rm i}$. 
Although the Mulliken analysis can not predict exactly the atomic charges quantitatively,
the sign of atomic charge can be estimated correctly\cite{szabo82}. 
 From the Mulliken analysis, we see
that  boron atoms in ${\rm C}_{48}{\rm B}_{12}$ exist as electron donor 
while  carbon atoms act as electron acceptors. The calculated charge of boron atom
is consistent with the experimental result of the Smalley group\cite{guo91} that an electron-deficient site
was produced at the boron position on the cage.

\begin{center}
\epsfig{file=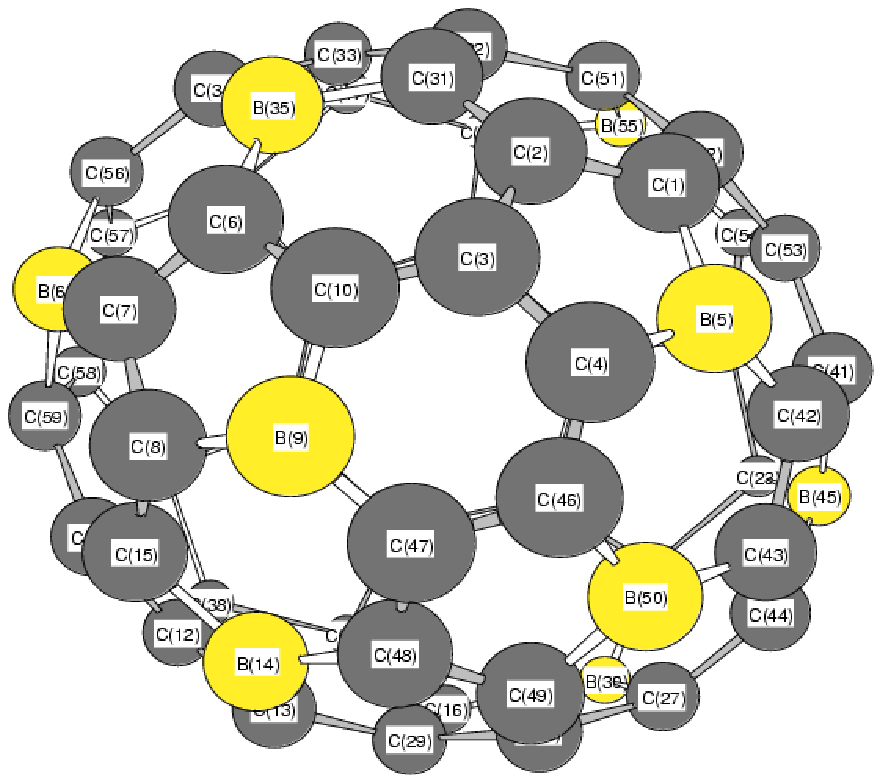,width=6.5cm,height=7cm}
\end{center}

{\bf FIG.1}: ${\rm C}_{48}{\rm B}_{12}$  geometric structure optimized with B3LYP/6-31G(d). The site numbers
\{5, 9, 14, 21, 26, 30, 35, 39, 45, 50, 55, 60\}  are for boron atoms and the others for carbon atoms.

\noindent
{\bf Table I:} B3LYP/6-31G(d) calculations of radius (${\rm R}_{i}$, in nm) and
 net Mulliken charge (${\rm Q}_{i}$, in $e$, where  $1\ e=1.6\times 10^{-19}$ C) for
${\rm C}_{48}{\rm B}_{12}$.
 
\begin{center}
\begin{tabular}{cccc}\hline\hline
 Site Number $\{ n_{i} \}$ & Atom
 & ${\rm R}_{i}$ & ${\rm Q}_{i}$  \\ \hline
\{1, 13, 16, 31, 38, 51\} &C &0.35198  &-0.0385     \\
\{2, 12, 29, 32, 37, 52\} &C &0.34652  &-0.0079    \\
\{3, 11, 28, 33, 36, 53\} &C &0.35701  &-0.0036     \\
\{4, 15, 27, 34, 40, 54\} &C &0.37033  &-0.0712     \\
\{5, 14, 30, 35, 39, 55\} &B &0.36454  &0.1637     \\
\{6, 18, 24, 42, 48, 58\} &C &0.37116  &-0.0333     \\
\{7, 19, 23, 43, 47, 57\} &C &0.38035  &-0.0376    \\
\{8, 20, 22, 44, 46, 56\} &C&0.38252  &-0.0826    \\
\{9, 21, 26, 45, 50, 60\} &B &0.37953  &0.1871      \\
\{10, 17, 25, 41, 49, 59\}&C&0.37079  &-0.0761     \\ \hline
\end{tabular}
\end{center}

\

The optimized carbon-carbon (CC) and boron-carbon (BC) bond lengths in
${\rm C}_{48}{\rm B}_{12}$ are listed in Table II. We find that ${\rm C}_{48}{\rm B}_{12}$
has 6 unique BC bond lengths in the range of 0.15376 nm to 0.15899 nm  and 9 unique
CC bond lengths in the range of 0.13861 nm to 0.15014 nm.  For ${\rm C}_{60}$ molecule, the
double  C=C bond and single C-C bond lengths are 0.13949 nm and 0.14539 nm, respectively,
in excellent agreement with experiment\cite{burgi92}.

\begin{center}{\bf III. Electric (HYPER)POLARIZABILITY}\end{center}

The static dipole polarizability (SDP) for heterofullerene C$_{48}$B$_{12}$  is  presented in
Table III. The B3LYP results were  obtained by using the Gaussian
98 program \cite{gaussian,nist},  while the LDA (local density approximation)
results were calculated by using the Amsterdam Density Functional (ADF)
program \cite{nist,ADF:2002,jcc_22_931}. The  SDPs for C$_{48}$N$_{12}$  and C$_{60}$ listed 
in Table III are taken from our recent work \cite{xie02c}. For the B3LYP calculations, we use
the valence-split basis set 6-31G(d) including the polarization functions for  boron and  carbon 
atoms. The ADF program uses basis sets of Slater functions. In this work, we use 
a triple zeta valence plus polarization (TZP) augmented with field-induced
polarization (FIP) functions of Zeiss et al.~\cite{Zeiss:79}. This basis set,
TZP++ ([6s4p2d1f] for carbon and boron atoms), was previously used for 
calculating the second-order hyperpolarizability $\gamma$ of C$_{60}$ and its
derivatives (for example, C$_{58}$N$_{2}$, C$_{58}$B$_{2}$ and C$_{58}$BN) and has
been shown to produce reasonable (hyper)polarizabilities even with
its small size~\cite{Jensen:02}. Here, we  only make a comparison
between C$_{48}$B$_{12}$, C$_{48}$N$_{12}$ and C$_{60}$. A comparison of
C$_{60}$ with other theoretical and experimental results can be found in
review chapters\cite{book4a,book4b} and recent work \cite{xie02c,Jensen:02}.

From Table III, we see that the LDA results are about
20\% larger than the corresponding B3LYP ones. This is expected since the basis set in
the LDA calculation  is larger and the LDA method, in general, predicts a larger polarizability
than the B3LYP method \cite{jcp_109_10180}. Nevertheless, both B3LYP and LDA predict
the same trends. The mean polarizability of C$_{48}$B$_{12}$  in the LDA (B3LYP)  is
about 12\% ( 15\%) larger than that of C$_{60}$.

The first hyperpolarizability $\beta$ of C$_{48}$B$_{12}$ is zero due to  
inversion symmetry. The static second-order hyperpolarizability,
$\gamma$, for C$_{48}$B$_{12}$, C$_{48}$N$_{12}$ and C$_{60}$ are presented in
Table IV. The $\gamma$ values for C$_{48}$N$_{12}$ and C$_{60}$ listed in Table IV 
are taken from Ref.~\cite{xie02c}, and the comparison will only be made between these three molecules.
For the calculations of the second-order hyperpolarizability, we use time-dependent DFT (TD-DFT)
method as described in Ref.~\cite{Jensen:02,vanGisbergen:97}, i.e.
finite-field differentiation of the analytically calculated first-order
hyperpolarizability. For all TD-DFT calculations, we used the RESPONSE
code~\cite{nist,vanGisbergen:99} implemented in the ADF program.~\cite{nist,ADF:2002,jcc_22_931}.
 
We find for all components of the second-order hyperpolarizability for both C$_{48}$B$_{12}$ and
C$_{48}$N$_{12}$ a larger value than for C$_{60}$. All  $\gamma$ components for C$_{48}$B$_{12}$ are
also larger than the corresponding components for C$_{48}$N$_{12}$ except for the 
$\gamma_{zzzz}$. This gives an average second-order hyperpolarizability, 
$\overline{\gamma}$ of C$_{48}$B$_{12}$, which is about 180\% larger than that of C$_{60}$.
In contrast, the $\overline{\gamma}$ value for C$_{48}$N$_{12}$ is about 55 \% larger than that of
C$_{60}$. The increase in the second-order hyperpolarizability is much
larger for C$_{48}$B$_{12}$ than for C$_{48}$N$_{12}$ especially considering that there is no increase in
volume. It has been experimentally shown that ${\rm C}_{60}$ is a good optical limiter
\cite{book4b} because of its larger $\gamma$ value.  Our present results imply that
heterofullerene C$_{48}$B$_{12}$ can  compete with ${\rm C}_{60}$ as an even better
optical limiter because of its enhanced third-order optical nonlinearity.

\end{multicols}

\noindent
{\bf Table II:} B3LYP/6-31G(d) calculation of CC and BC bond lengths (in nm)  in
molecule ${\rm C}_{48}{\rm B}_{12}$.
\begin{center}
\begin{tabular}{cccccccccc}\hline\hline
Bond&   Site Number Pairs for Bonding &  Bond Length \\ \hline
 CC & (1,  2) (12,  13) (16, 29) (31, 32) (37, 38) (51, 52) &0.14261 \\
 BC & (1,  5) (13,  14) (16, 30) (31, 35) (38, 39) (51, 55) &0.15376 \\
 CC & (1, 52) (2 ,  31) (12, 38) (13, 29) (16, 37) (32, 51) &0.14078 \\
 CC & (2,  3) (11,  12) (28, 29) (32, 33) (36, 37) (52, 53) &0.15014  \\
 CC & (3,  4) (11,  15) (27, 28) (33, 34) (36, 40) (53, 54) &0.14871 \\
 CC & (3, 10) (11,  59) (17, 33) (25, 36) (28, 49) (41, 53) &0.13861 \\
 BC & (4,  5) (14,  15) (27, 30) (39, 40) (34, 35) (54, 55) &0.15667 \\
 CC & (4, 46) ( 8,  15) (20, 40) (22, 54) (27, 44) (34, 56) &0.13867  \\
 BC & (5, 42) ( 6,  35) (14, 48) (18, 55) (24, 30) (39, 58) &0.15576 \\
 CC & (6, 10) (17,  18) (24, 25) (58, 59) (41, 42) (48, 49) &0.14682 \\
 CC & (6,  7) (18,  19) (23, 24) (42, 43) (47, 48) (57, 58) &0.14080 \\
 BC & (7, 60) (9,   47) (19, 26) (21, 57) (23, 45) (43, 50) &0.15594  \\
 CC & (7,  8) (19,  20) (22, 23) (43, 44) (46, 47) (56, 57) &0.14570  \\
 BC & (8,  9) (20,  21) (22, 26) (44, 45) (46, 50) (56, 60) &0.15899  \\
 BC & (9, 10) (17,  21) (25, 26) (41, 45) (49, 50) (59, 60) &0.15563  \\ \hline
\end{tabular}
\end{center}

\noindent
{\bf Table III:} Static dipole polarizability ($\alpha$, in nm$^3$)
for C$_{48}$B$_{12}$, C$_{48}$N$_{12}$ and C$_{60}$ calculated with B3LYP/6-31G(d)
 and LDA/TZP++. The symmetry relation for C$_{60}$ gives
$\alpha_{xx}=\alpha_{yy}=\alpha_{zz}$ and for C$_{48}$B$_{12}$ and C$_{48}$N$_{12}$
is $\alpha_{xx}=\alpha_{yy}$.
 
\begin{center}
 
\begin{tabular}{lcccccc}\hline\hline
 &\multicolumn{2}{c}{B3LYP/6-31G(d)}& \ \ \ \  &\multicolumn{2}{c}{LDA/TZP++} & \\
\cline{2-3}\cline{5-6}
Molecule&$\alpha_{xx}$&$\alpha_{zz}$ & &$\alpha_{xx}$&$\alpha_{zz}$ & Ref.\\\hline
C$_{60}$         & 0.0695 &0.0695    & &0.0847 &0.0847 & \cite{xie02c}\\
C$_{48}$N$_{12}$ & 0.0666 &0.0675    & &0.0793 &0.0815 & \cite{xie02c}\\
C$_{48}$B$_{12}$ & 0.0822 &0.0804    & &0.0958 &0.0939 & this work \\ \hline
\end{tabular}
\end{center}
 
\

\noindent
{\bf Table IV:} The static second-order hyperpolarizabilities ($\gamma$, in a.u., with
1 a.u. = 6.235378$\times$10$^{-65}$ C$^4$m$^4$J$^{-3}$)  for
C$_{48}$B$_{12}$ , C$_{48}$N$_{12}$ and C$_{60}$ calculated  by using LDA and a TZP++ basis set.
The average second-order  hyperpolarizability is given by
$\overline{\gamma} = {1 \over 15}\sum_{i,j}(\gamma_{iijj}+\gamma_{ijij}+\gamma_{ijji})$. The symmetry
relations of the molecule gives
$\gamma_{xxxx}=\gamma_{yyyy}$, $\gamma_{xxzz} = \gamma_{yyzz}$ and $\gamma_{zzxx}=\gamma_{zzyy}$.
 
\begin{center}
 
\begin{tabular}{lcccccccccc}\hline\hline
Molecule &$\gamma_{xxxx}$&$\gamma_{xxyy}$&$\gamma_{zzzz}$&$\gamma_{xxzz}$&$\gamma_{zzxx}$&$\overline{\gamma}$ & Ref.\\\hline
C$_{60}$         &137950 &45983  &137950 &45983  &45983  &137950 &\cite{xie02c}\\
C$_{48}$N$_{12}$ &188780 &62880  &232970 &85120  &84790  &215222 &\cite{xie02c}\\
C$_{48}$B$_{12}$ &470190 &156840 &214300 &116800 &118090 &387628 & this work \\ \hline
\end{tabular}
\end{center}

\begin{multicols}{2}

\begin{center}{\bf IV. IR AND RAMAN SPECTRA}\end{center}

Using the Gausian 98 program\cite{gaussian,nist}, we first optimize the geometry of ${\rm C}_{48}{\rm B}_{12}$ and 
${\rm C}_{60}$ with the B3LYP method and 3-21G basis set. Then, we calculate the vibrational frequencies of ${\rm C}_{48}{\rm B}_{12}$ 
and ${\rm C}_{60}$ with the same method and basis set.  Our results for ${\rm C}_{60}$ are in agreement with experiment\cite{ir1,rm1}. 
C$_{60}$ has totally 46 vibrational modes\cite{book1}. Since ${\rm C}_{48}{\rm B}_{12}$  has lower symmetry ($C_{i}$) than ${\rm C}_{60}$, we find 
 174 independent vibrational modes for ${\rm C}_{48}{\rm B}_{12}$: 87 non-degenerate IR-active modes 
with $a_{u}$ symmetry  and 87 non-degenerate Raman-active modes with $a_{g}$ symmetry.  
 
We also calculate  IR intensities $I_{IR}$  and Raman scattering activities $\Omega_{raman}$ at 
the corresponding vibrational frequencies  for both ${\rm C}_{60}$ 
and heterofullerene ${\rm C}_{48}{\rm B}_{12}$. The results are shown in Fig.2 and Fig.3. 
Since experimental IR and Raman spectroscopic 
data do not directly indicate the specific type of nuclear motion producing each spectroscopic peak, we 
do not give here the normal mode displacement  for the vibrational frequencies. 

For ${\rm C}_{60}$, we note that its IR spectrum is very simple. Namely, it
is composed of 4 IR spectroscopic signals with $t_{1u}$ symmetry.
The IR intensities for ${\rm C}_{60}$
calculated with B3LYP/3-2G agree reasonably with the {\sl in situ} high-resolution FTIR spectrum of a ${\rm C}_{60}$ film
measured by Onoe and Takeuchi\cite{irc60exp}. However, the IR spectrum of
${\rm C}_{48}{\rm B}_{12}$ is not so simple, exhibiting 87 IR spectroscopic signals, with the stronger IR spectroscopic signals
mainly in the high-frequency region. Here we discuss several vibrational bands
 in the IR intensity of ${\rm C}_{48}{\rm B}_{12}$ to be compared with future experimental identification: (i) the strongest
IR spectroscopic signal with IR intensity of 229912 m/mol is determined at the high frequency 1356 cm$^{-1}$; (ii) three strong
modes are observed at   frequencies $\nu$ = 783  cm$^{-1}$, 1031 cm$^{-1}$ and 1546 cm$^{-1}$ with IR intensities of
85828 m/mol, 119322 m/mol and 114878 m/mol, respectively; (iii) five intermediate modes appear at
 frequencies $\nu$ = 730  cm$^{-1}$, 1018 cm$^{-1}$, 1106 cm$^{-1}$, 1217  cm$^{-1}$ and 1303  cm$^{-1}$
with IR intensities of 58303  m/mol, 69950 m/mol, 55094  m/mol, 69160  m/mol and  53856  m/mol, respectively.

\begin{center}
\epsfig{file=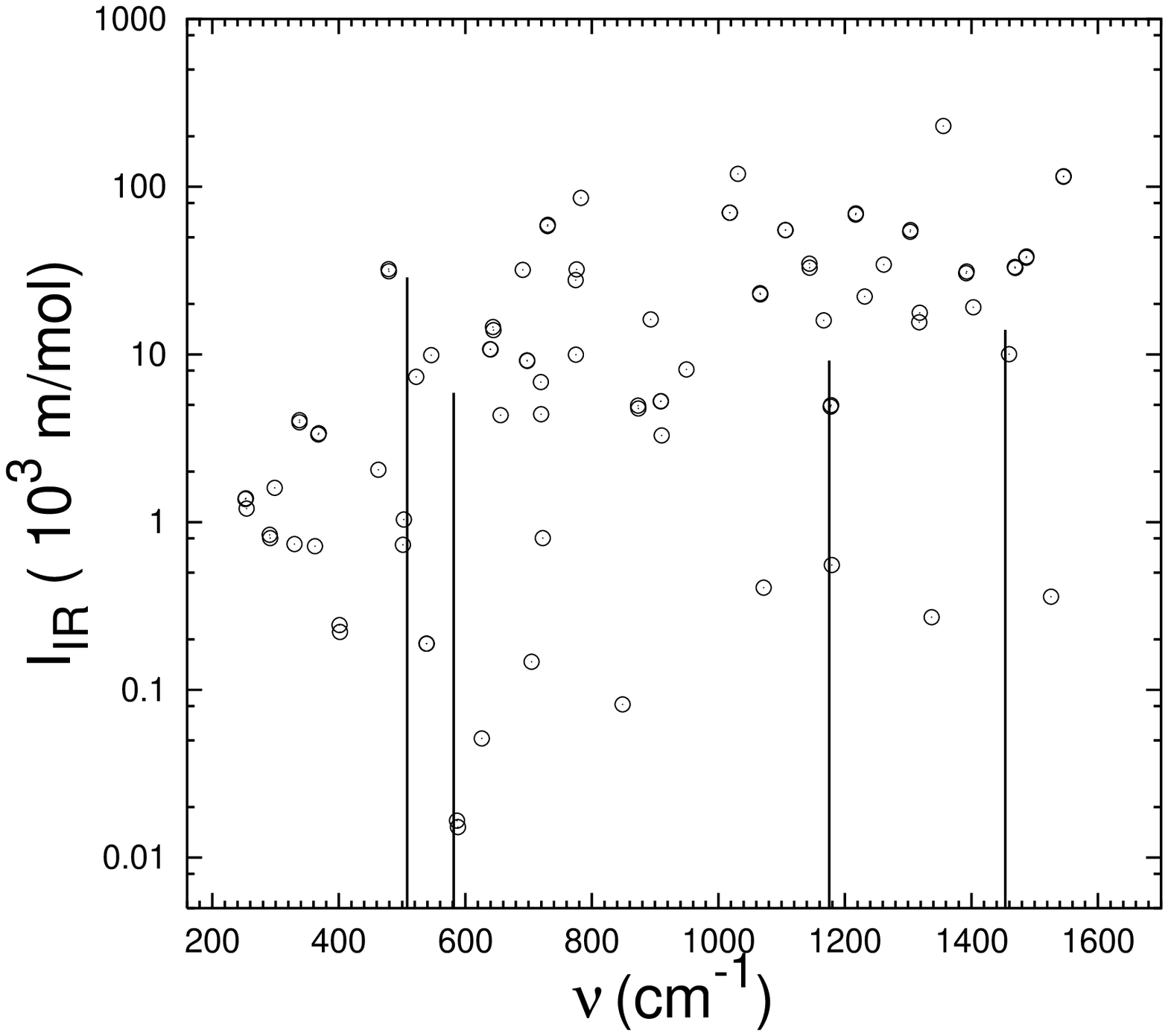,width=6cm,height=5cm}
\end{center}

{\bf Fig.2}: B3LYP/3-21G calculation of IR-active vibrational
frequencies ($\nu$, in cm$^{-1}$) and IR intensities ($I_{IR}$, in $10^{3}\ {\rm m/mole}$)
of ${\rm C}_{48}{\rm B}_{12}$ (open circles) and ${\rm C}_{60}$ (solid lines).

\begin{center}
\epsfig{file=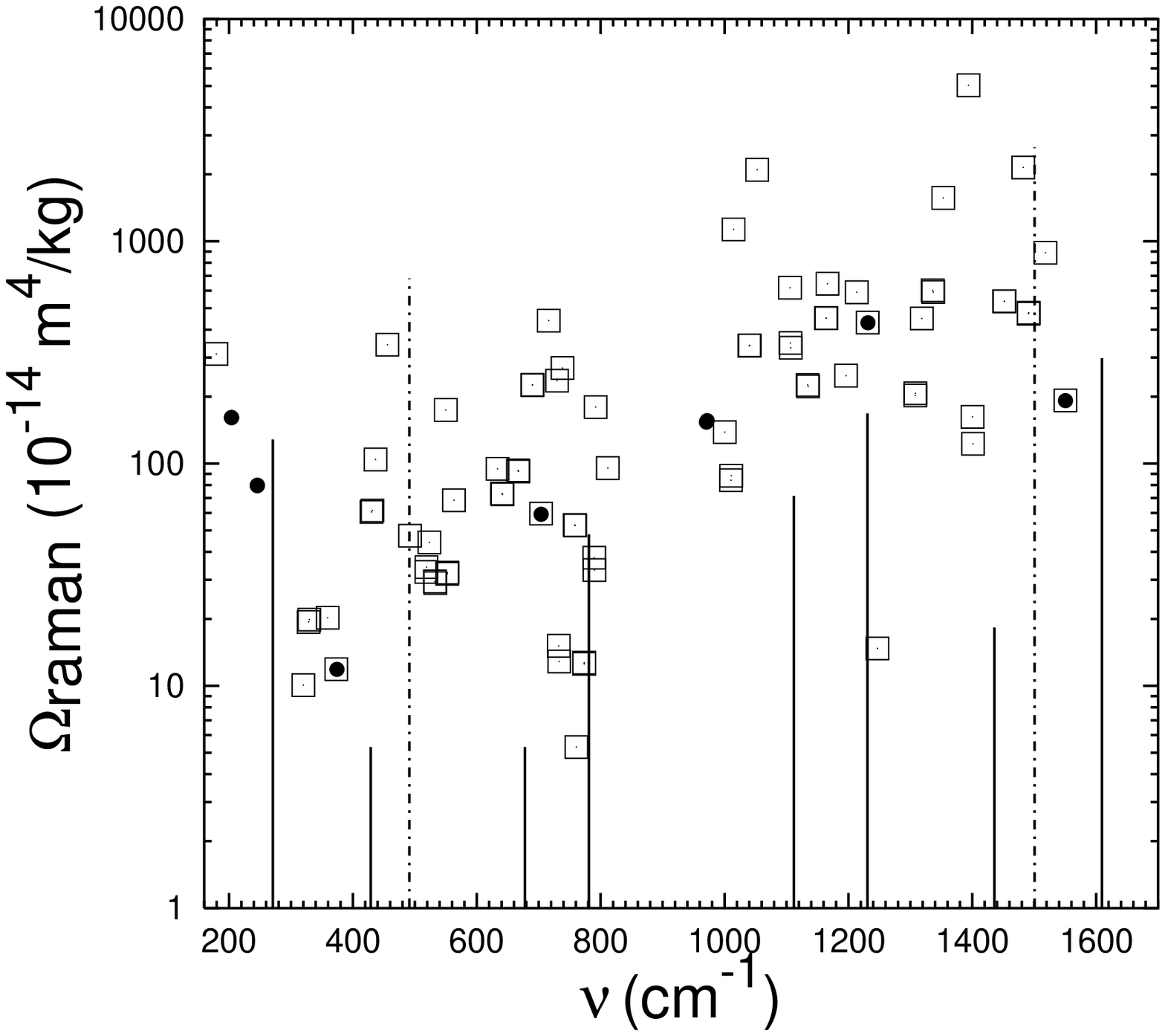,width=6cm,height=5cm}
\end{center}

{\bf Fig.3}:  B3LYP/3-21G calculations of Raman-active frequencies ($\nu$, in cm$^{-1}$) and Raman
scattering activities ( $\Omega_{raman}$, in  $10^{-14} {\rm m}^{4}/{\rm kg}$)
of ${\rm C}_{48}{\rm B}_{12}$. The solid and dot-dashed lines are the unpolarized and polarized
Raman spectral lines of ${\rm C}_{60}$, respectively. Filled circles and open squares are
non-degenerate unpolarized and polarized Raman-active modes, respectively.

\

As shown in Fig.3,  ${\rm C}_{60}$  has two non-degenerate polarized Raman
spectroscopic signals with  ${\rm a}_{\rm g}$ symmetry and 8 fivefold-degenerate unpolarized ones
with ${\rm h}_{\rm g}$ symmetry. The strongest Raman spectroscopic signals in ${\rm C}_{60}$ are
the two ${\rm a}_{\rm g}$ modes. The calculated results for ${\rm C}_{60}$ are 
in excellent agreement with experiment\cite{rm1}. In contrast,
for ${\rm C}_{48}{\rm B}_{12}$,  we observe 10 non-degenerate  unpolarized  and 77 non-degenerate
polarized Raman spectroscopic signals with $a_{g}$ symmetry. The Raman spectrum separates into 
  high-frequency (1000 ${\rm cm}^{-1}$ to 1700 ${\rm cm}^{-1}$) and low-frequency
(200 ${\rm cm}^{-1}$ to 900 ${\rm cm}^{-1}$) regions,  similar to those of ${\rm C}_{60}$.  The 
strong Raman spectroscopic signals in ${\rm C}_{48}{\rm B}_{12}$ are the non-degenerate polarized  
modes. Three strong Raman bands of ${\rm C}_{48}{\rm B}_{12}$  are: (i) The
strongest Raman spectroscopic signal in ${\rm C}_{48}{\rm B}_{12}$ appears at high frequency 
$\nu$ = 1394 cm$^{-1}$ with $\Omega_{raman} = 5035\times 10^{-14}$ m$^{4}$/kg; (ii) four strong 
Raman spectroscopic signals  are located at $\nu$ =  1014 cm$^{-1}$, 1053 cm$^{-1}$, 1353 cm$^{-1}$ and 
 1482  cm$^{-1}$ with $\Omega_{raman} = 1130\times 10^{-14}$ m$^{4}$/kg, 
 $2094\times 10^{-14}$ m$^{4}$/kg, $1568\times 10^{-14}$ m$^{4}$/kg and  
$2149\times 10^{-14}$ m$^{4}$/kg, respectively;  
(iii) five intermediate strong Raman spectroscopic signals are observed at $\nu$ = 1105 cm$^{-1}$, 
1166 cm$^{-1}$, 1214 cm$^{-1}$, 1336 cm$^{-1}$ and 1518 cm$^{-1}$ with  $\Omega_{raman} = 618\times 10^{-14}$ m$^{4}$/kg, 
$644\times 10^{-14}$ m$^{4}$/kg, $590\times 10^{-14}$ m$^{4}$/kg, $601\times 10^{-14}$ m$^{4}$/kg, 
$537\times 10^{-14}$ m$^{4}$/kg and $887\times 10^{-14}$ m$^{4}$/kg, respectively.

\begin{center}{\bf V.  SECOND-ORDER MAGNETIC RESPONSE}\end{center}  

There are a number of theoretical methods for calculating the second-order magnetic response 
properties of molecules. In this paper, we use both the gauge-including-atomic-orbital (GIAO) 
method and the continuous-set-of-gauge-transformation (CSGT) procedure\cite{jrc96}, which is 
implemented in the Gaussian 98 program\cite{gaussian,nist}, to predict the NMR shielding tensors 
$\sigma$ of ${\rm C}_{48}{\rm B}_{12}$. 
In high-resolution  NMR, the isotropic part $\sigma_{iso}$  of
$\sigma$ is measured by taking the average of $\sigma$ with respect to the orientation 
to the magnetic field, i.e.,  $\sigma_{iso} =
(\sigma_{xx}+\sigma_{yy}+\sigma_{zz})/3$,
where $\sigma_{xx}$, $\sigma_{yy}$ and $\sigma_{zz}$ are the principal
axis values of  $\sigma$. The results calculated by using B3LYP hybrid DFT and restricted 
Hartree Fock (RHF) theory are summarized in Table V. We find
that  ${\rm C}_{48}{\rm B}_{12}$ has 8 $^{13}{\rm C}$ and 2 $^{11}{\rm B}$ 
NMR  spectral signals, indicative of the 10 unique sites  in the 
${\rm C}_{48}{\rm B}_{12}$ structure. In contrast,
${\rm C}_{60}$ has only one $^{13}{\rm C}$ NMR spectral signal, for example, 
 with $\sigma_{iso}=50.5\ {\rm ppm}$  (parts per million) and 54.7 ppm obtained with 
B3LYP/6-31G(d):GIAO and RHF/6-31G(d):GIAO, respectively. The $^{13}$C NMR chemical shift
($\delta=\sigma_{iso}^{TMS}-\sigma_{iso}^{sample}$) with  respect to
the reference tetramethysilane (TMS) for ${\rm C}_{60}$ is, for example, 
 $\delta = 133.3 (135.7)$  ppm  for B3LYP/6-31G(d):GIAO (B3LYP/6-31G(d):CSGT), about 9 (7) ppm difference from 
experiment ($\delta = 142.7$ ppm\cite{nmrexp}), but $\delta=140.4 (141.7)$ ppm for RHF/6-31G(d):GIAO 
(RHF/6-31G(d):CSGT) which is in good agreement with  experiment. The results for ${\rm C}_{60}$ 
show that the DFT  method does not provide systematically better NMR results than RHF. This is 
due to the fact that no current functionals include a magnetic field dependence\cite{jrc96}. 
For ${\rm C}_{60}$, the CSGT procedure provides better NMR results than the 
GIAO procedure, but takes more CPU time (see Table V) than the GIAO procedure when 
compared to experiment.

\begin{center}{\bf VI. SUMMARY}\end{center}
 
In summary, we have performed  first-principles calculations of bonding, Mulliken charges, dipole
polarizability, hyperpolarizability, vibrational frequencies, IR intensities,  Raman
scattering activities and second-order magnetic response properties of
heterofullerene ${\rm C}_{48}{\rm B}_{12}$.
Eighty-seven independent IR-active and 87 independent Raman-active vibrational modes for
${\rm C}_{48}{\rm B}_{12}$ are assigned. Eight $^{13}$C and two $^{11}$B NMR spectral lines
for ${\rm C}_{48}{\rm B}_{12}$ are characterized. Compared to ${\rm C}_{60}$ and ${\rm C}_{48}{\rm N}_{12}$,
 ${\rm C}_{48}{\rm B}_{12}$ exhibits enhanced third-order optical nonlinearity, which
implies potential applications of ${\rm C}_{48}{\rm B}_{12}$ in photonics and optical limiting.

\end{multicols}
 
\noindent
{\bf Table V:}   B3LYP/6-31G(d) and RHF/6-31G(d) calculations of the absolute
isotropy, $\sigma_{iso}$ in  ppm (parts per million),  of the nuclear magnetic shielding tensor
$\sigma$  for atoms  in ${\rm C}_{48}{\rm B}_{12}$, ${\rm C}_{60}$ and tetramethysilane (TMS)
 found by using  both GIAO and CSGT methods. The CPU times (in hours) for ${\rm C}_{60}$ cases are shown.

\begin{center}
\begin{tabular}{cccccccccccccc}\hline\hline
 &\multicolumn{10}{c}{${\rm C}_{48}{\rm B}_{12}$} & TMS &\multicolumn{2}{c}{ ${\rm C}_{60}$} \\
\cline{2-11}\cline{13-14} {\small Theoretical Method}  &  $^{13}$C[1] &  $^{13}$C[2] &  $^{13}$C[3] &  $^{13}$C[4] &  $^{11}$B[5]
&  $^{13}$C[6] & $^{13}$C[7] &  $^{13}$C[8] &  $^{11}$B[9] &  $^{13}$C[10] & $^{13}$C  &  $^{13}$C & CPU \\ \hline
{\small B3LYP/6-31G(d):GIAO} &  42.9 & 19.1 & 25.6 & 27.4 & 65.2 & 2.7 & 17.1 & 32.3 & 73.6 & 37.6 &183.8 & 50.5 & 26.6    \\
{\small B3LYP/6-31G(d):CSGT} & 39.3 & 16.0 & 21.9 & 23.7 & 60.3 &0.8 & 13.9 & 28.8 & 69.1 & 34.0   & 181.6 & 45.9 & 28.5  \\
  \\
{\small RHF/6-31G(d):GIAO}   & 59.7 & 21.2 & 41.4 & 52.2 & 73.1 & 24.6 & 36.8 & 37.2 & 75.2 & 56.4 & 195.1 & 54.7 & 13.5   \\
{\small RHF/6-31G(d):CSGT} & 55.0 & 17.5 & 36.9 & 47.1 & 68.4 & 20.8 & 31.7 & 33.0 & 71.0 & 51.9 & 190.6 & 48.9 & 18.5  \\ \hline
\end{tabular}
\end{center}

\begin{multicols}{2}

\section*{Acknowledgements}
We thank Dr. Denis A. Lehane and Dr. Hartmut Schmider for their 
technical help. One of us (R. H. X.) thanks the HPCVL at Queen's 
University for the use of its parallel supercomputing facilities.  L.J. 
gratefully acknowledges the Danish Research Training Council for financial 
support. V.H.S. acknowledges support from the Natural Science and Engineering 
Research Council of Canada (NSERC).

\end{multicols}
 

\begin{references}

\bibitem{kroto85}H.W. Kroto, J.R. Heath, S.C. Obrien, R.E. Curl, and R.E. Smalley, 
Nature (London) {\bf 318}, 162 (1985).


\bibitem{krat90}W. Kr\"{a}tschmer, L.D. Lamb, K. Fostiropoulos, and D.R. Huffman, Nature (London) {\bf 347}, 354 (1990).

\bibitem{kroto93}H.W. Kroto, J.E. Fischer, and D.E. Cox, {\sl The Fullerenes} 
(Pergamon, Oxford, 1993).
 
\bibitem{book1} M.S. Dresselhaus, G. Dresselhaus, and P. C. Eklund, {\sl 
Science of Fullerenes and Carbon Nanotubes} (Academic Press, New York, 1996).

 
\bibitem{book4a}R.H. Xie,   {\sl Chapter 6: Nonlinear Optical Properties of Fullerenes and Carbon Nanotubes }, 
in: {\sl Handbook of Advanced Electronic and Photonic Materials and Devices, Vol. 9: Nonlinear Optical Materials}, 
H. S. Nalwa (Ed.) (Academic Press, New York, 2000), pp.267-307.

\bibitem{book4b}R.H. Xie, Q. Rao, and L. Jensen, {\sl Nonlinear Optics of Fullerenes and Carbon Nanotubes}, 
in: {\sl Encyclopedia of Nanoscience and Nanotechnology}, H. S. Nalwa (Ed.) (American Scientific Publisher, California, 2003).

 
\bibitem{hummelen95} J.C. Hummelen, B. Knight, J. Pavlovich, R.Gonz\'{a}lez, and F. Wudl, Science {\bf 269}, 1554 (1995).

\bibitem{guo91} T. Guo, C.M. Jin, and R.E. Smalley, J. Phys. Chem. {\bf 95}, 4948 (1991).

\bibitem{hultman01} L. Hultman, S. Stafstr\"{o}m, Z. Czig\'{a}ny, 
J. Neidhardt, N. Hellgren, I. F. Brunell, K. Suenaga, and 
C. Cooliex,  Phys. Rev. Lett. {\bf 87}, 225503 (2001).

\bibitem{stafstrom} S. Stafstr\"{o}m, L. Hultman, and N. Hellgren, 
Chem. Phys. Lett. {\bf 340}, 227 (2001).

\bibitem{xie02a} R. H. Xie, G. W. Bryant, and V. H. Smith, Jr., 
Chem. Phys. Lett. {\bf 368}, 486 (2003).

\bibitem{mana02}M.R. Mana, D.W. Sprehn, and H.A. Ichord, J. Am. Chem. Soc. 
{\bf 124}, 13990 (2002).

\bibitem{xie02b} R.H. Xie, G.W. Bryant, J. Zhao,  V.H. Smith, Jr.,
A. Di Carlo, and A. Pecchia,  Phys. Rev, Lett., accepted for publication.  

\bibitem{gaussian} Gaussian 98, Revision A.9,
 M.J. Frisch, G.W. Trucks, H.B. Schlegel, G.E. Scuseria,
 M.A. Robb, J.R. Cheeseman, V.G. Zakrzewski, J.A. Montgomery, Jr.,
 R.E. Stratmann, J.C. Burant, S. Dapprich, J.M. Millam,
 A.D. Daniels, K.N. Kudin, M.C. Strain, O. Farkas, J. Tomasi,
 V. Barone, M. Cossi, R. Cammi, B. Mennucci, C. Pomelli, C. Adamo,
 S. Clifford, J. Ochterski, G.A. Petersson, P.Y. Ayala, Q. Cui,
 K. Morokuma, D.K. Malick, A.D. Rabuck, K. Raghavachari,
 J.B. Foresman, J. Cioslowski, J.V. Ortiz, A.G. Baboul,
 B.B. Stefanov, G. Liu, A. Liashenko, P. Piskorz, I. Komaromi,
 R. Gomperts, R.L. Martin, D.J. Fox, T. Keith, M.A. Al-Laham,
 C.Y. Peng, A. Nanayakkara, M. Challacombe, P.M. W. Gill,
 B. Johnson, W. Chen, M.W. Wong, J.L. Andres, C. Gonzalez,
 M. Head-Gordon, E.S. Replogle, and J.A. Pople,
 Gaussian, Inc., Pittsburgh PA, 1998.

\bibitem{nist} Use of this software does not constitute an endorsement
or certification by NIST.

\bibitem{becke93} A.D. Becke, J. Chem. Phys.  {\bf  98}, 5648 (1993).
 
\bibitem{slater74} J.C. Slater, Phys. Rev. {\bf 81}, 385  (1951).
 
\bibitem{becke88} A.D. Becke, Phys. Rev. A {\bf 38}, 3088 (1988).
 
\bibitem{vosko80} S.H. Vosko, L. Wilk, and M. Nusair, Can. J. Phys. {\bf 58}, 1200 (1980).
 
\bibitem{lyp88} C. Lee, W. Yang,  and R.G. Parr, Phys. Rev. B {\bf 37}, 785 (1988).


\bibitem{burgi92} H. B. B\"{u}rgi, E. Blanc, D. Schwarzenbach,
S. Liu, Y. Lu, M. M. Kappes, and J. A. Ibers, Angew.
Chem. Int. Ed. Engl. {\bf 41}, 640 (1992).

 
\bibitem{szabo82} A. Szabo and N. S. Ostlund, {\sl Modern Quantum Chemistry}
(Macmillan, New York, 1982).


\bibitem{ADF:2002}
{ADF 2002.01}., {T}heoretical {C}hemistry {V}rije  {U}niversiteit, {A}msterdam (2002).
 
\bibitem{jcc_22_931}
G. te Velde, F.M.  Bickelhaupt, E.J. Baerends, C. Fonseca Guerra, S.J.A. van
  Gisbergen, J.G. Snijders, and T. Ziegler.  J. Comp. Chem. {\bf 22}, 931 (2001).
 
 
\bibitem{xie02c} R.H. Xie, G.W. Bryant, L. Jensen, J. Zhao, and V.H. Smith, Jr.,
J. Chem. Phys. {\bf 118}, in May,  2003. 
 
\bibitem{Zeiss:79}
G.D. Zeiss, W.R. Scott, N. Suzuki, and D.P. Chong, Mol. Phys. {\bf 37},  1543 (1979).
 
\bibitem{Jensen:02}
L. Jensen, P.Th. van Duijnen, J.G. Snijders, and D.P. Chong, Chem. Phys. Lett. {\bf 359}, 524 (2002).

\bibitem{jcp_109_10180}
D.J. Tozer and N.C. Handy, J. Chem. Phys. {\bf 109}, 10180 (1998).

\bibitem{vanGisbergen:97}
S.J.A. van Gisbergen, J.G. Snijders, and E.J. Baerends, Phys. Rev. Lett. {\bf 78}, 3097 (1997) .

\bibitem{vanGisbergen:99}
S.J.A. van Gisbergen, J.G. Snijders, and E.J. Baerends, Comput. Phys. Commun. 
{\bf 118}, 119 (1999).

\bibitem{ir1} W. Kr\"{a}tschmer, K. Fostiropoulos, and D. R. Huffman,
Chem. Phys. Lett. {\bf 170}, 167 (1990).
 
\bibitem{rm1} K. Lynch, C. Tanke, F. Menzel, W. Brockner, P. Scharff
and E. Stumpp,  J. Phys. Chem. {\bf 99}, 7985 (1995).


\bibitem{irc60exp}J. Onoe and K. Takeuchi, Phys. Rev. B {\bf 54}, 6167 (1996).

\bibitem{jrc96}J.R. Cheeseman, M.J. Frisch, G.W. Trucks, and T.A. Keith, 
J. Chem. Phys. {\bf 104}, 5497 (1996).

\bibitem{nmrexp}R. Taylor, J.P. Hare, A.K. Adul-Sada and H.W. Kroto, 
J. Chem. Soc. Chem. Commun., 1423 (1990).

\end{references}
\end{document}